\documentclass[journal]{IEEEtran}
\usepackage{cite}

\ifCLASSINFOpdf
\else
\fi
\usepackage{graphicx}

\usepackage{amsmath}
\usepackage{mathrsfs}
\usepackage[dvipsnames]{xcolor}
\usepackage[colorlinks]{hyperref}
\allowdisplaybreaks
\begin{document}
%
\title{On the Probability of Erasure for MIMO-OFDM}
%
%
%

\author{K. Vasudevan, ~\IEEEmembership{Senior Member,~IEEE,}
        A. Phani Kumar Reddy, 
        Gyanesh Kumar Pathak
        and~Shivani Singh 
\thanks{The authors are with the Department
of Electrical Engineering, Indian Institute of Technology, 
Kanpur-208016, India
e-mail: \{vasu, phani, pathak, shivanis\}@iitk.ac.in.}
}

%
%

\markboth{SSID, Feb~2020}%
{Vasudevan \MakeLowercase{\textit{et al.}}: Probability of Erasure for MIMO-OFDM}
%



\maketitle

\begin{abstract}
Detecting the presence of a valid signal is an important task 
of a telecommunication receiver. When the receiver is unable to detect 
the presence of a valid signal, due to noise and fading, it is
referred to as an erasure. This work deals with the probability
of erasure computation for orthogonal frequency division multiplexed
(OFDM) signals used by multiple input multiple output (MIMO) systems.
The theoretical results are validated by computer simulations.
OFDM is widely used in present day wireless communication systems due
to its ability to mitigate intersymbol interference (ISI) caused by 
frequency selective fading channels. MIMO systems offer the advantage
of spatial multiplexing, resulting in increased bit-rate, which is
the main requirement of the recent wireless standards like 5G and
beyond.
\end{abstract}


%
\IEEEpeerreviewmaketitle

\section{Introduction}
%
%
%
%
\IEEEPARstart{A}{s} we move towards higher bit-rates and millimeter-wave
frequencies, the physical layer of a telecommunication system needs
to maintain reliable communication between the transmitter and receiver.
This implies that
\begin{enumerate}
 \item The operating average signal-to-noise ratio per bit  (or
       $E_b/N_0$) must be as close to 0 dB as possible
       \cite{KV_OpSigPJ2019,73ddc0ea-7d42-4fdd-969d-da08c8e4d0c0}. This ensures
       that the transmit power is kept as low as possible, which enhances the
       battery life of a mobile handset. It needs to be again mentioned that
       the operating $E_b/N_0$ of wireless telecommunication systems e.g. a
       mobile handset, is not specified, only the received signal strength is
       mentioned \cite{Vasudevan2015}. If $E_b/N_0$ was such an important
       parameter in the universities, why was it dropped by the industry?
 \item For a given $E_b/N_0$, the bit-error-rate must be minimized. In fact,
       it has been shown in \cite{KV_OpSigPJ2019} that it
       is possible to achieve error-free transmission as long as $E_b/N_0$
       is greater than $-1.6$ dB even for fading channels.
 \item The transmission bandwidth should be minimized, by use of
       pulse shaping \cite{Vasu_Book10}. This allows spectrum sharing among a
       large number of users.
\end{enumerate}

In order to achieve the aforementioned targets, the receiver must be
optimally designed, incorporating the best signal processing algorithms. It
is well known that the optimum receiver is a coherent receiver, that has perfect
knowledge of the carrier frequency and phase, timing phase and the channel
impulse response. One of the methods to obtain a near-coherent receiver is to
train it with a known preamble, before the commencement of data (information)
transmission. Now, the question here is: how does the receiver know that
the preamble has been transmitted? Moreover, in the presence of noise and
fading, the receiver may not be able to detect the presence of a preamble.
This condition is referred to as an erasure. The subject of this work is to
compute the probability of erasure for a multiple input multiple output (MIMO)
orthogonal frequency multiplexed (OFDM) system, since it is expected to have
applications in 5G and beyond. This work has not been done earlier.

The probability of erasure was simulated for single input single output (SISO)
OFDM in \cite{6663392,c7888430-cbc2-4c14-87bb-b52780478d85}. Subsequent works on
single input multiple output (SIMO) OFDM \cite{Vasudevan2015} and MIMO OFDM
\cite{Vasu_ICWMC2016,Vasu_IARIA2017,Vasu_intech:2019,
d4bbbdf0-7468-4727-9ebe-76d5e6160b64}
also dealt with the probability of erasure, since the receiver was trained with
a known preamble. However, the simulation results for the probability of
erasure were not published.

This work is orgnized as follows. Section~\ref{Sec:Sys_Model} presents the
system model. The probability of erasure is derived in 
Section~\ref{Sec:Prob_Erasure}. Computer simulation results are given in
Section~\ref{Sec:Results}. Conclusions and future work are mentioned in
Section~\ref{Sec:Conclude}.

\begin{figure*}[tbhp]
\begin{center}
\input{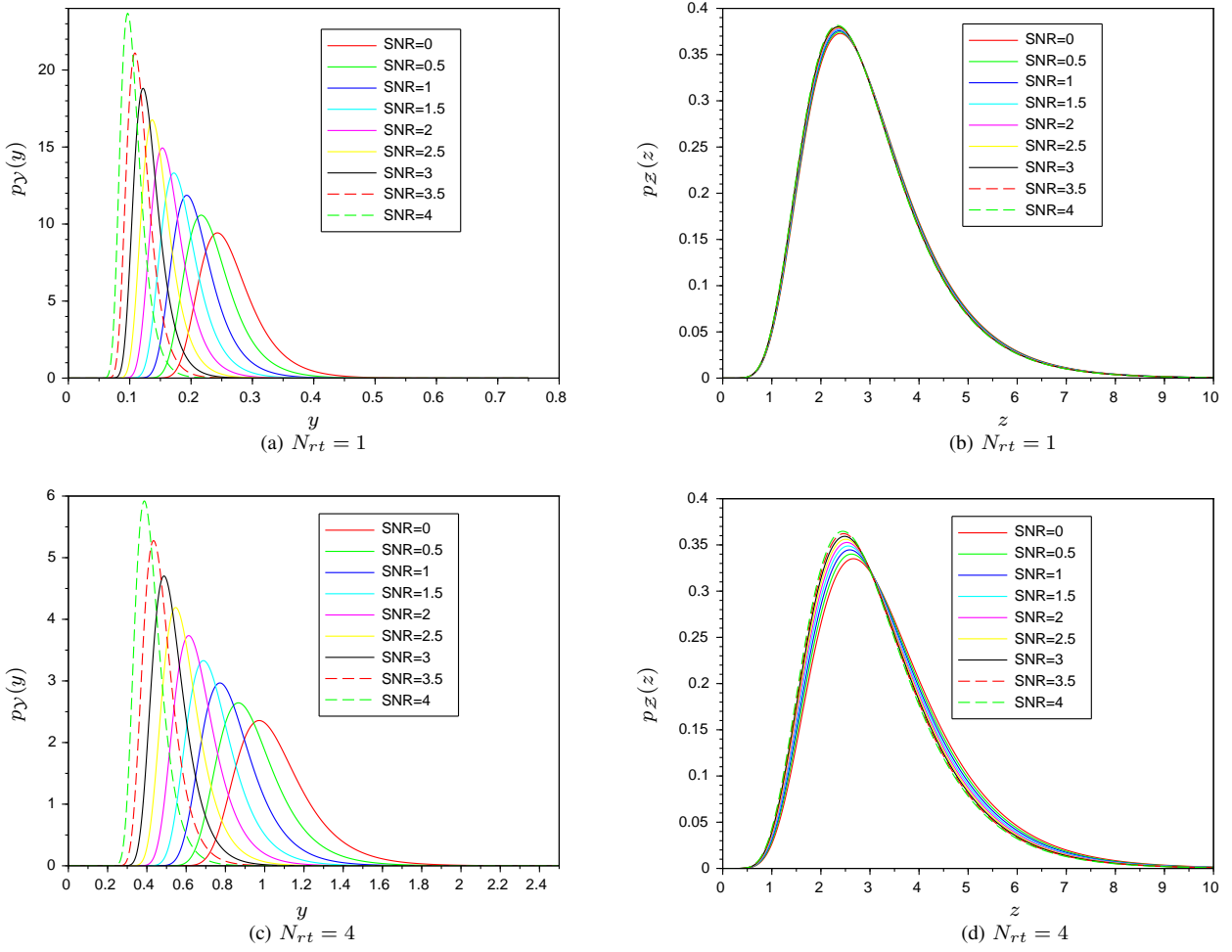}
\caption{PDF of $Y$ and $Z$ for various SNRs, re-transmissions and $N=4$ 
         antennas.}
\label{Fig:PDF_YZ_NT4}
\end{center}
\end{figure*}
\begin{figure*}[tbhp]
\begin{center}
\input{pdf_yz_nt8.pstex_t}
\caption{PDF of $Y$ and $Z$ for various SNRs, re-transmissions and $N=8$
         antennas.}
\label{Fig:PDF_YZ_NT8}
\end{center}
\end{figure*}
\section{System Model}
\label{Sec:Sys_Model}
Consider an $N\times N$ MIMO system (see Figure~1 of \cite{Vasu_intech:2019}). 
The channel is assumed to
be quasi-static (time-invariant over one re-transmission and varies randomly
over re-transmissions), frequency selective fading. The channel impulse
response between transmit antenna $n_t$ and receive antenna $n_r$ for the
$k^{th}$ re-transmission at time $n$ is denoted by
$\tilde h_{k,\, n,\, n_r,\, n_t}$ and is characterized by (E3) in
\cite{Vasu_intech:2019}. The subscripts in the channel impulse response
are integers that lie in the range $1\le k\le N_{rt}$,
$0\le n < L_h$ and $1 \le n_r,\, n_t \le N$. Note that the number of
re-transmissions is denoted by $N_{rt}$, the length of the
channel impulse response is $L_h$ and the number of antennas at the
transmitter and receiver is $N$.
Clearly, there are $N^2 N_{rt}$ channel impulse responses to be estimated.
This is accomplished by using the frame structure given in Figure~3(a) of
\cite{Vasu_intech:2019}. Recall that in the preamble phase, only one transmit
antenna is active at an time, whereas in the data phase all transmit antennas
are simultaneously active. The preamble symbols are drawn from a QPSK
constellation and uncoded. The received signal
$\tilde r_{k,\, n,\, n_r,\, n_t,\, p}$ during the preamble phase
is given by (E8) of \cite{Vasu_intech:2019}. We assume that there is no 
frequency offset, that is $\omega_0=0$ in (E8) of \cite{Vasu_intech:2019}. The
effect of non-zero frequency
offset is studied in the computer simulations. At the correct timing instant
(perhaps a circular shift of $\tilde r_{k,\, n,\, n_r,\, n_t,\, p}$ may be
necessary), the $L_p$-point FFT of $\tilde r_{k,\, n,\, n_r,\, n_t,\, p}$
yields (E13) in \cite{Vasu_intech:2019}, which is repeated here for convenience
($0\le i < L_p$, $i$ is an integer):
\begin{equation}
\label{Eq:Pap13_Eq1}
\tilde R_{k,\, i,\, n_r,\, n_t,\, p} =
                           \tilde H_{k,\, i,\, n_r,\, n_t}
                            S_{1,\, i} +
                           \tilde W_{k,\, i,\, n_r,\, n_t,\, p}.
\end{equation}
The subscript ``$p$'' in the above equation denotes the preamble phase. Now
\begin{align}
\label{Eq:Pap13_Eq2}
\frac{\tilde R_{k,\, i,\, n_r,\, n_t,\, p} S_{1,\, i}^*}{E_s}
                     & = \tilde H_{k,\, i,\, n_r,\, n_t}
                         \frac{\left|S_{1,\, i}\right|^2}{E_s} \nonumber  \\
                     &   \qquad +
                         \frac{\tilde W_{k,\, i,\, n_r,\, n_t,\, p}
                          S_{1,\, i}^*}{E_s}                   \nonumber  \\
                     & = \tilde H_{k,\, i,\, n_r,\, n_t} +
                         \frac{\tilde W_{k,\, i,\, n_r,\, n_t,\, p}
                          S_{1,\, i}^*}{E_s}
\end{align}
since (see (E10) of \cite{Vasu_intech:2019})
\begin{equation}
\label{Eq:Pap13_Eq3}
\left|
 S_{1,\, i}
\right|^2 = E_s
\end{equation}
denotes the energy of the preamble. The $L_p$-point inverse fast Fourier
transform (IFFT) of (\ref{Eq:Pap13_Eq2}) yields (for $0\le n < L_p$, $n$ is an
integer)
\begin{equation}
\label{Eq:Pap13_Eq4}
\tilde r_{1,\, k,\, n,\, n_r,\, n_t,\, p} =
\tilde h_{k,\, n,\, n_r,\, n_t} +
\tilde w_{1,\, k,\, n,\, n_r,\, n_t,\, p}
\end{equation}
where the second term on the right hand side of (\ref{Eq:Pap13_Eq4}) is the
$L_p$-point IFFT of the second term on the right hand side of 
(\ref{Eq:Pap13_Eq2}).
From (\ref{Eq:Pap13_Eq4}) it is clear that $0\le n < L_h$ contains the
channel coefficients plus noise, whereas $L_h\le n < L_p$ contains only noise.
Typically
\begin{equation}
\label{Eq:Pap13_Eq4_0}
L_h \ll L_p.
\end{equation}
Moreover
\begin{equation}
\label{Eq:Pap13_Eq4_1}
\frac{1}{2}
 E
\left[
\left|
\tilde w_{1,\, k,\, n,\, n_r,\, n_t,\, p}
\right|^2
\right] = \frac{\sigma^2_w}{E_s}
\end{equation}
where $\sigma^2_w$ is the one-dimensional variance of
$\tilde w_{k,\, n,\, n_r,\, n_t,\, p}$
in (E8) of \cite{Vasu_intech:2019}. For a given $k$, $n_r$ and $n_t$ let
\begin{align}
\label{Eq:Pap13_Eq5}
\tilde Z_n & = \tilde h_{k,\, n,\, n_r,\, n_t} +
               \tilde w_{1,\, k,\, n,\, n_r,\, n_t,\, p}
               \qquad \mbox{for $0\le n < L_h$}              \nonumber  \\
\tilde Y_n & = \tilde w_{1,\, k,\, n,\, n_r,\, n_t,\, p}
               \qquad \mbox{for $L_h \le n < L_p$}.
\end{align}
Note that $\tilde Z_n$ and $\tilde Y_n$ are complex Gaussian random variables
with zero-mean. Due to independence between the channel coefficients and 
noise
\begin{align}
\label{Eq:Pap13_Eq6}
\frac{1}{2}
 E
\left[
\left|
\tilde Z_n
\right|^2
\right] & = \sigma^2_f + \frac{\sigma^2_w}{E_s}       \nonumber  \\
        &   \stackrel{\Delta}{=} \sigma^2_Z           \nonumber  \\
\frac{1}{2}
 E
\left[
\left|
\tilde Y_n
\right|^2
\right] & = \frac{\sigma^2_w}{E_s}                    \nonumber  \\
        &   \stackrel{\Delta}{=} \sigma^2_Y
\end{align}
where $\sigma^2_f$ is the one-dimensional variance of
$\tilde h_{k,\, n,\, n_r,\, n_t}$ (see (E3) of \cite{Vasu_intech:2019}).
An erasure occurs when
\begin{equation}
\label{Eq:Pap13_Eq7}
\max_n
\left|
\tilde Z_n
\right|^2 <
\max_n
\left|
\tilde Y_n
\right|^2
\end{equation}
\begin{figure*}[tbhp]
\begin{center}
\input{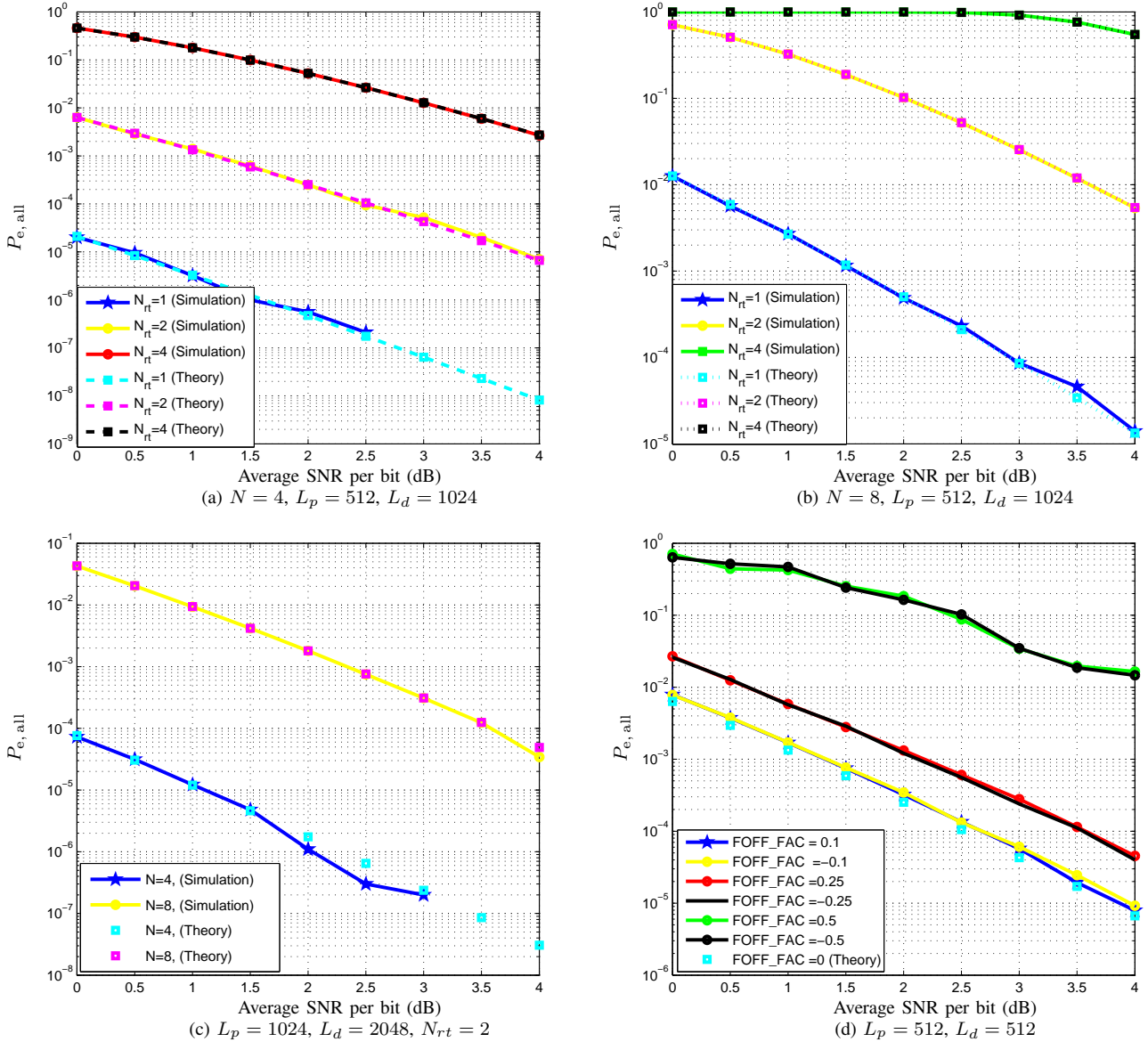}
\caption{Simulation results.}
\label{Fig:Pap13_Results}
\end{center}
\end{figure*}
\section{The Probability of Erasure}
\label{Sec:Prob_Erasure}
The probability density function (pdf) of
$\mathscr{Z}_n=\left|\tilde Z_n\right|^2$ and
$\mathscr{Y}_n=\left|\tilde Y_n\right|^2$ is given by
(for $0 \le z,\, y < \infty$) \cite{Proakis95}
\begin{align}
\label{Eq:Pap13_Eq8}
p_{\mathscr{Z}_n}(z) & = \frac{1}{2\sigma^2_Z}
                         \mathrm{e}^{-z/(2\sigma^2_Z)}     \nonumber  \\
p_{\mathscr{Y}_n}(y) & = \frac{1}{2\sigma^2_Y}
                         \mathrm{e}^{-y/(2\sigma^2_Y)}
\end{align}
where we have assumed that $\tilde Z_n$ and $\tilde Y_n$ are wide sense
stationary, so that their pdfs are independent of the time index $n$. Let
\begin{align}
\label{Eq:Pap13_Eq9}
\mathcal{Z} & = \max_n \mathscr{Z}_n                \nonumber  \\
\mathcal{Y} & = \max_n \mathscr{Y}_n.
\end{align}
Since $\mathscr{Z}_n$ and $\mathscr{Y}_n$ are independent over $n$, the
cumulative distribution function (cdf) of $\mathcal{Z}$ and $\mathcal{Y}$ is
\cite{Papoulis91,Vasu_AC_PS}
\begin{align}
\label{Eq:Pap13_Eq10}
 P
\left(
\mathcal{Z}\le z
\right) & = \left(
             P
            \left(
            \mathscr{Z}_n<z
            \right)
            \right)^{L_h}                            \nonumber  \\
        & = \left(
             1 - \mathrm{e}^{-z/(2\sigma^2_Z)}
            \right)^{L_h}                            \nonumber  \\
 P
\left(
\mathcal{Y}\le y
\right) & = \left(
             P
            \left(
            \mathscr{Y}_n<y
            \right)
            \right)^{L_p-L_h}                        \nonumber  \\
        & = \left(
             1 - \mathrm{e}^{-y/(2\sigma^2_Y)}
            \right)^{L_p-L_h}.
\end{align}
Therefore, the pdf of $\mathcal{Z}$ and $\mathcal{Y}$ are
(for $0 \le z,\, y<\infty$)
\begin{align}
\label{Eq:Pap13_Eq11}
p_{\mathcal{Z}}(z)
        & = \frac{d}{dz}
             P
            \left(
            \mathcal{Z}\le z
            \right)                                    \nonumber  \\
        & = \frac{L_h}{2\sigma^2_Z}
            \mathrm{e}^{-z/(2\sigma^2_Z)}
            \left(
             1 - \mathrm{e}^{-z/(2\sigma^2_Z)}
            \right)^{L_h-1}                            \nonumber  \\
p_{\mathcal{Y}}(y)
        & = \frac{d}{dy}
             P
            \left(
            \mathcal{Y}\le y
            \right)                                    \nonumber  \\
        & = \frac{L_p-L_h}{2\sigma^2_Y}
            \mathrm{e}^{-y/(2\sigma^2_Y)}
            \left(
             1 - \mathrm{e}^{-y/(2\sigma^2_Y)}
            \right)^{L_p-L_h-1}.                       \nonumber  \\
\end{align}
For a given $k$, $n_r$, $n_t$, and from (\ref{Eq:Pap13_Eq7}) there is no
erasure when
\begin{align}
\label{Eq:Pap13_Eq12}
P_{\mathrm{ne},\, 1} & = \int_{z=0}^{\infty}
                          P(\mathcal{Y}\le z|z)
                          p_{\mathcal{Z}}(z)\, dz     \nonumber  \\
                     & = \int_{z=0}^{\infty}
                         \int_{y=0}^{z}
                          p_{\mathcal{Y}}(y)\, dy
                         \,
                          p_{\mathcal{Z}}(z)\, dz
\end{align}
where the subscript ``ne'' denotes ``no erasure'' and ``1'' denotes one
particular value of $k$, $n_r$ and $n_t$.
Now
\begin{align}
\label{Eq:Pap13_Eq13}
I(z) & =  P(\mathcal{Y}\le z|z)                      \nonumber  \\
     & = \int_{y=0}^{z}
          p_{\mathcal{Y}}(y)\, dy                    \nonumber  \\
     & = \frac{L_p-L_h}{2\sigma^2_Y}
         \int_{y=0}^{z}
         \sum_{l=0}^{L_p-L_h-1}
          (-1)^l
         \binom{L_p-L_h-1}{l}                        \nonumber  \\
     &   \qquad
         \times
         \mathrm{e}^{-(l+1)y/(2\sigma^2_Y)}          \nonumber  \\
     & = (L_p-L_h)
         \sum_{l=0}^{L_p-L_h-1}
         \frac{(-1)^l}{l+1}
         \binom{L_p-L_h-1}{l}                        \nonumber  \\
     &   \qquad
         \times
         \left(
          1 -
         \mathrm{e}^{-(l+1)z/(2\sigma^2_Y)}
         \right).
\end{align}
Let
\begin{equation}
\label{Eq:Pap13_Eq14}
A_l = \frac{L_p-L_h}{l+1}
      (-1)^l
      \binom{L_p-L_h-1}{l}.
\end{equation}
Then
\begin{equation}
\label{Eq:Pap13_Eq15}
I(z) = \sum_{l=0}^{L_p-L_h-1}
        A_l
       \left(
        1 -
       \mathrm{e}^{-(l+1)z/(2\sigma^2_Y)}
       \right).
\end{equation}
Therefore
\begin{align}
\label{Eq:Pap13_Eq16}
P_{\mathrm{ne},\, 1} & = \int_{z=0}^{\infty}
                          I(z)p_{\mathcal{Z}}(z)\, dz         \nonumber  \\
                     & = \int_{z=0}^{\infty}
                         \sum_{l=0}^{L_p-L_h-1}
                          A_l
                         \left(
                          1 -
                         \mathrm{e}^{-(l+1)z/(2\sigma^2_Y)}
                         \right)
                          p_{\mathcal{Z}}(z)\, dz             \nonumber  \\
                     & = \frac{L_h}{2\sigma^2_Z}
                         \sum_{l=0}^{L_p-L_h-1}
                          A_l
                         \left(
                          1 -
                         \int_{z=0}^{\infty}
                         \mathrm{e}^{-(l+1)z/(2\sigma^2_Y)}
                         \right.                              \nonumber  \\
                     &   \qquad \times
                         \left.
                         \sum_{a=0}^{L_h-1}
                         (-1)^a
                         \binom{L_h-1}{a}
                         \mathrm{e}^{-(a+1)z/(2\sigma^2_Z)}
                         \right)\, dz                         \nonumber  \\
                     & = \frac{L_h}{2\sigma^2_Z}
                         \sum_{l=0}^{L_p-L_h-1}
                          A_l
                         \left(
                          1 -
                         \sum_{a=0}^{L_h-1}
                         (-1)^a
                         \binom{L_h-1}{a}
                         \right.                              \nonumber  \\
                     &   \qquad \times
                         \left.
                         \int_{z=0}^{\infty}
                         \mathrm{e}^{-(l+1)z/(2\sigma^2_Y)}
                         \mathrm{e}^{-(a+1)z/(2\sigma^2_Z)}
                         \right)\, dz                         \nonumber  \\
                     & = \sum_{l=0}^{L_p-L_h-1}
                          A_l'
                         \left(
                          1 -
                         \sum_{a=0}^{L_h-1}
                         (-1)^a
                         \binom{L_h-1}{a}
                         \frac{1}{B_l+C_a}
                         \right)                              \nonumber  \\
\end{align}
where
\begin{align}
\label{Eq:Pap13_Eq17}
A_l' & = \frac{L_h}{2\sigma^2_Z} A_l         \nonumber  \\
B_l  & = \frac{l+1}{2\sigma^2_Y}             \nonumber  \\
C_a  & = \frac{a+1}{2\sigma^2_Z}.
\end{align}
Since the channel and noise are independent across $k$, $n_r$ and $n_t$ the
probability of no erasure for all $k$, $n_r$ and $n_t$ is
\begin{equation}
\label{Eq:Pap13_Eq18}
P_{\mathrm{ne},\,\mathrm{all}} =
\left(
 P_{\mathrm{ne},\, 1}
\right)^{N_{rt}N^2}.
\end{equation}
Finally, the probability of erasure for all $k$, $n_r$, $n_t$ is
\begin{equation}
\label{Eq:Pap13_Eq19}
P_{\mathrm{e},\,\mathrm{all}} = 1-P_{\mathrm{ne},\,\mathrm{all}}.
\end{equation}
While it appears that (\ref{Eq:Pap13_Eq19}) is the closed form expression for
the probability of erasure over all re-transmissions ($k$), receive antennas
($n_r$) and transmit antennas ($n_t$), the main problem lies in the
computation of $A_l$ in (\ref{Eq:Pap13_Eq14}). The reason is that
$\binom{L_p-L_h-1}{l}$ takes very large values (results in overflow) for
preamble length $L_p=512$ and channel length $L_h=10$. In order to alleviate 
this problem, we need to resort to numerical integration techniques. Thus
(\ref{Eq:Pap13_Eq12}) is approximated by
\begin{equation}
\label{Eq:Pap13_Eq20}
P_{\mathrm{ne},\, 1}     \approx
                         \sum_{i_0=0}^{N_{\mathcal{Z}}}
                         \sum_{i_1=0}^{i_0 \Delta z}
                          p_{\mathcal{Y}}(i_1 \Delta y)\, \Delta y
                         \,
                          p_{\mathcal{Z}}(i_0 \Delta z)\, \Delta z
\end{equation}
where
\begin{align}
\label{Eq:Pap13_Eq21}
\Delta y        & = \Delta z      \nonumber  \\
                & =  10^{-3}      \nonumber  \\
N_{\mathcal{Z}} & =  10^4         \nonumber  \\
i_0\Delta z     & =  z            \nonumber  \\
i_1\Delta y     & =  y.
\end{align}
Note that
\begin{equation}
\label{Eq:Pap13_Eq22}
N_{\mathcal{Z}} \Delta z =  10.
\end{equation}
In other words, it is sufficient to take the maximum value of $z$ in
the upper limit of the integral over $z$ in (\ref{Eq:Pap13_Eq12}) to be equal
to 10.
This is also clear from the plots of the pdf of $\mathcal{Z}$ in
Figures~\ref{Fig:PDF_YZ_NT4} and \ref{Fig:PDF_YZ_NT8} as a function of
the average SNR per bit, the number of re-transmissions $N_{rt}$ and the
number of antennas $N$. The average SNR per bit is defined in (E21) of
\cite{Vasu_intech:2019}, and is repeated here for convenience
\begin{equation}
\label{Eq:Pap13_Eq23}
\mbox{SNR}_{\mathrm{av},\, b,\, p} = \frac{4L_h\sigma^2_f N N_{rt}}
                                          {L_d\sigma^2_w}.
\end{equation}
We find that in all cases, the pdf of $\mathcal{Z}$ goes to zero for $z>10$,
which justifies the relation in (\ref{Eq:Pap13_Eq22}).
It can also be seen from Figures~\ref{Fig:PDF_YZ_NT4} and \ref{Fig:PDF_YZ_NT8}
that the pdf of $\mathcal{Z}$ is relatively insensitive to variations
in $N$ and $N_{rt}$, whereas the pdf of $\mathcal{Y}$ varies widely with
$N$ and $N_{rt}$.
\section{Results}
\label{Sec:Results}
The probability of erasure results are presented in
Figure~\ref{Fig:Pap13_Results}. The results for four transmit and receive
antennas ($N=4$) is shown in Figure~\ref{Fig:Pap13_Results}(a). We see that the
theoretical and simulated curves nearly coincide, which demonstrates the
accuracy of our prediction. Moreover, the probability of
erasure increases with increasing re-transmissions. The probability of erasure
with eight transmit and receive antennas ($N=8$) is given in
Figure~\ref{Fig:Pap13_Results}(b). Clearly, the probability of erasure is three
orders of magnitude higher than $N=4$, for the same number of
re-transmissions $N_{rt}$. Next, in Figure~\ref{Fig:Pap13_Results}(c), the
length of the preamble and data is doubled compared to 
Figure~\ref{Fig:Pap13_Results}(a) and (b), that is $L_p=1024$, $L_d=2048$
and the number of re-transmissions is fixed to $N_{rt}=2$. We see that the
probability of erasure reduces by two orders of magnitude. In other words,
increasing the length of the preamble ($L_p$) reduces the probability of 
erasure. Note that, when the preamble length is increased, the data length
($L_d$) also needs to be increased, to keep the throughput fixed (see
Table 1 of \cite{Vasu_intech:2019}). Finally in 
Figure~\ref{Fig:Pap13_Results}(d), we plot the probability of erasure in the
presence of frequency offset ($\omega_0$). The received signal model in the 
presence of frequency offset is given by (E8) of \cite{Vasu_intech:2019},
before the FFT operation at the receiver. We have taken
\begin{equation}
\label{Eq:Pap13_Eq24}
\omega_0 = \verb!FOFF_FAC! \times \frac{2\pi}{L_p}
\end{equation}
where $2\pi/L_p$ denotes the subcarrier spacing of the preamble. We see that
the probability of erasure increases with increasing \verb!FOFF_FAC!.
\section{Conclusion}
\label{Sec:Conclude}
In this work, we have derived the probability of erasure (defined as the
probability of not detecting an OFDM frame when it is transmitted) for
MIMO-OFDM systems. It is assumed that the frequency offset has been
accurately estimated and cancelled. Simulation results are also presented in
the presence of frequency offset, and it is 
seen that the probability of erasure increases with increasing frequency
offset. Future work could be to derive the
probability of erasure in the presence of small frequency offsets, say for
$\verb!FOFF_FAC!<0.25$. The other interesting area could be to compute
$\binom{a}{b}$ for large values of $a$ and $b$.

\bibliographystyle{IEEEtran}
\bibliography{mybib,mybib1,mybib2,mybib3,mybib4}

\begin{thebibliography}{10}
\providecommand{\url}[1]{#1}
\csname url@samestyle\endcsname
\providecommand{\newblock}{\relax}
\providecommand{\bibinfo}[2]{#2}
\providecommand{\BIBentrySTDinterwordspacing}{\spaceskip=0pt\relax}
\providecommand{\BIBentryALTinterwordstretchfactor}{4}
\providecommand{\BIBentryALTinterwordspacing}{\spaceskip=\fontdimen2\font plus
\BIBentryALTinterwordstretchfactor\fontdimen3\font minus
  \fontdimen4\font\relax}
\providecommand{\BIBforeignlanguage}[2]{{%
\expandafter\ifx\csname l@#1\endcsname\relax
\typeout{** WARNING: IEEEtran.bst: No hyphenation pattern has been}%
\typeout{** loaded for the language `#1'. Using the pattern for}%
\typeout{** the default language instead.}%
\else
\language=\csname l@#1\endcsname
\fi
#2}}
\providecommand{\BIBdecl}{\relax}
\BIBdecl

\bibitem{KV_OpSigPJ2019}
K.~Vasudevan, K.~Madhu, and S.~Singh, ``{Data Detection in Single User Massive
  MIMO Using Re-Transmissions},'' \emph{The Open Signal Processing Journal},
  vol.~6, pp. 15--26, Mar. 2019.

\bibitem{73ddc0ea-7d42-4fdd-969d-da08c8e4d0c0}
------, ``Scilab code for data detection in single user massive mimo using
  re-transmissions,'' \url{https://www.codeocean.com/}, 6 2019.

\bibitem{Vasudevan2015}
\BIBentryALTinterwordspacing
K.~Vasudevan, ``Coherent detection of turbo-coded ofdm signals transmitted
  through frequency selective rayleigh fading channels with receiver diversity
  and increased throughput,'' \emph{Wireless Personal Communications}, vol.~82,
  no.~3, pp. 1623--1642, 2015. [Online]. Available:
  \url{http://dx.doi.org/10.1007/s11277-015-2303-8}
\BIBentrySTDinterwordspacing

\bibitem{Vasu_Book10}
------, \emph{{Digital Communications and Signal Processing, Second edition
  (CDROM included)}}.\hskip 1em plus 0.5em minus 0.4em\relax Universities Press
  (India), Hyderabad, www.universitiespress.com, 2010.

\bibitem{6663392}
------, ``Coherent detection of turbo coded ofdm signals transmitted through
  frequency selective rayleigh fading channels,'' in \emph{Signal Processing,
  Computing and Control (ISPCC), 2013 IEEE International Conference on}, Sept.
  2013, pp. 1--6.

\bibitem{c7888430-cbc2-4c14-87bb-b52780478d85}
------, ``Scilab code for coherent detection of turbo coded ofdm signals
  transmitted through frequency selective rayleigh fading channels,''
  \url{https://www.codeocean.com/}, 7 2019.

\bibitem{Vasu_ICWMC2016}
------, ``Coherent turbo coded mimo ofdm,'' in \emph{ICWMC 2016, The 12th
  International Conference on Wireless and Mobile Communications}, Nov. 2016,
  pp. 91--99,
  \href{http://www.thinkmind.org/download.php?articleid=icwmc_2016_6_30_20101}{[Online]}.

\bibitem{Vasu_IARIA2017}
------, ``{Near Capacity Signaling over Fading Channels using Coherent Turbo
  Coded OFDM and Massive MIMO},'' \emph{International Journal On Advances in
  Telecommunications}, vol.~10, no. 1 \& 2, pp. 22--37, 2017,
  {\href{www.thinkmind.org/download.php?articleid=tele_v10_n12_2017_3}{[Online]}}.

\bibitem{Vasu_intech:2019}
K.~Vasudevan, S.~Singh, and A.~P.~K. Reddy, ``Coherent receiver for turbo coded
  single-user massive mimo-ofdm with retransmissions,'' in \emph{Multiplexing},
  S.~Mohammady, Ed.\hskip 1em plus 0.5em minus 0.4em\relax London: IntechOpen,
  2019, ch.~4, pp. 1--21.

\bibitem{d4bbbdf0-7468-4727-9ebe-76d5e6160b64}
------, ``Scilab code for coherent receiver for turbo coded single-user massive
  mimo-ofdm with retransmissions,'' \url{https://www.codeocean.com/}, 6 2019.

\bibitem{Proakis95}
J.~G. Proakis, \emph{{Digital Communications}}, 3rd~ed.\hskip 1em plus 0.5em
  minus 0.4em\relax McGraw Hill, 1995.

\bibitem{Papoulis91}
A.~Papoulis, \emph{{Probability, Random Variables and Stochastic Processes}},
  3rd~ed.\hskip 1em plus 0.5em minus 0.4em\relax McGraw-Hill, 1991.

\bibitem{Vasu_AC_PS}
K.~Vasudevan, \emph{{Analog Communications: Problems \& Solutions}}.\hskip 1em
  plus 0.5em minus 0.4em\relax Ane Books, 2018.

\end{thebibliography}

%








\end{document}